\documentclass[]{emulateapj}
\usepackage{natbib}
\usepackage{enumerate}

\newcommand{\mass}{\mathcal{M}}
\newcommand{\radius}{\mathcal{R}}

\newcommand{\SPHIGA}{{\sc{Sphiga}}}

\newcommand{\ve}[1]{\mathbf{#1}}

\newcommand{\refp}[1]{(\ref{#1})}
\newcommand{\Kepler}{\textit{Kepler}}
\newcommand{\s}[1]{\mathrm{#1}}
\newcommand{\icarus}{Icarus}

\begin{document}
\shorttitle{I. N-body Simulations}

\shortauthors{Meschiari}

\title{Circumbinary Planet Formation in the Kepler-16 system. I. N-body Simulations.}

\author{Stefano Meschiari\altaffilmark{1}}
\altaffiltext{1}{UCO/Lick Observatory, 
Department of Astronomy and Astrophysics, 
University of California at Santa Cruz,
Santa Cruz, CA 95064
}
\email{smeschia@ucolick.org}

\begin{abstract}
The recently discovered circumbinary planets (Kepler-16 b, Kepler 34-b, Kepler 35-b) represent the first direct evidence of the viability of planet formation in circumbinary orbits. We report on the results of $N$-body simulations investigating planetesimal accretion in the Kepler-16 b system, focusing on the range of impact velocities under the influence of both stars' gravitational perturbation and friction from a putative protoplanetary disk.  Our results show that planet formation might be effectively inhibited for a large range in semi-major axis (1.75  $\lesssim a_P \lesssim$ 4 AU), suggesting that the planetary core must have either migrated from outside 4 AU, or formed \emph{in situ} very close to its current location.
\end{abstract}

\section{Introduction}
The discovery of extrasolar planets around main-sequence stars is one of the major observational breakthroughs of the last decade.  The size of the planetary census, propelled by radial velocity (RV) surveys and dedicated missions such as \Kepler{}, has grown to include planetary systems where a variety of interesting dynamical interactions can be observed.  Such systems include 61 exoplanets discovered around stellar binaries\footnote{http://www.exoplanets.org, retrieved on February 14, 2012.} (including both planets orbiting one of the stellar companions and circumbinary planets). While for the majority of these planets the binarity of the system represents only a weak perturbation on the gravitational pull of the central star, a few single-planet systems have been detected in binaries with $a_\s{bin} \lesssim$ 30 AU (such as HD 41004, Gliese 86, HD196885 and $\gamma$ Cephei), with each planet in a circumstellar (``S-type'') orbit. Only one multiple system with $a_\s{bin} \lesssim 100$ AU has been found \citep[HD177830,][]{Meschiari11}.

The existence of these systems represents a major challenge to the current paradigm of planet formation. In fact, a number of simulations attempting to model the dynamics of the growth of planetary embryos from \textit{km}-sized planetesimals in presence of a binary companion have hit significant difficulties \citep[among others, ][]{Marzari00, Thebault02, Thebault04, Thebault06, Thebault11, Paardekooper08, Fragner11}. The most important parameter controlling planetesimal accretion is the mutual encounter velocity; indeed, runaway growth requires it to be less than the escape velocity for efficient accretion. The presence of the companion can stir up the relative velocity between planetesimals, interfering with runaway growth. Relative velocity is often excited beyond a fiducial threshold velocity at which all encounters are erosive, potentially slowing down planet formation or halting it altogether. Simulations taking into account a static background gas disk (representing an unperturbed protoplanetary disk at some point in time) initially posited that disk-planetesimals interaction induces a phasing of the orbits, making the environment more accretion-friendly \citep{Marzari00}. Nevertheless, if the protoplanets interact with the gas disk through aerodynamic drag alone, the phasing induced by the gas disk is clearly size-dependent, and protoplanets with different sizes will collide with large encounter speeds over the majority of the range in semi-major axis sampled \citep[][]{Thebault04}. Finally, a misalignment between the orbital plane of the binary and the gas disk can significantly affect the dynamics of the planetesimals. Small inclinations ($i_\s{B} < 10^\circ$) can  favor planetesimal accretion somewhat \citep{Xie09}. On the other hand, large
inclinations ($30^\circ < i_\s{B} < 50^\circ$) can significantly perturb the planetesimal disk, causing planetesimals to ``jump'' inwards and pile up into a smaller inner disk, where encounter velocities are more favorable to accretion \citep{Xie11}.

Most of the works in the literature have focused on observed or plausible circumstellar configurations (e.g. a planet orbiting one of the two stellar components), in light of the lack of direct evidence of the existence of circumbinary planets orbiting main-sequence stars, outside the realm of science fiction. Therefore, only a handful of articles have considered planet formation in circumbinary orbits \citep[``P-type''; e.g.,][]{Moriwaki04, Quintana06, Scholl07, Marzari08, Pierens08b, Pierens08}, and they lacked a reference observed configuration. 

Kepler 16-b \citep{Doyle11} is the first circumbinary planet that has been detected with \textit{Kepler}. The presence of a third object was first hinted through deviations of the timing of the stellar eclipses from a linear ephemeris. The definitive characterization as a planetary object came from transits on both star A (tertiary eclipse) and star B (quaternary eclipse). The planet was determined to be a Saturn-mass planet ($\mass_\s{P} \approx 0.33 \mass_\s{Jup}$) on a nearly circular 228-day orbit; long-term integrations have shown the planet to be stable, with an eccentricity oscillating between 0 and $\approx 0.08$. The binary stellar system is composed of two main-sequence stars in an eccentric 41-day orbit, with a mass of 0.69 and 0.2 $\mass_\s{\odot}$ (mass ratio $\mu \approx 0.2$), respectively. The close coplanarity of the binary and planetary orbital planes suggests that the three bodies were formed in a common disk. This was bolstered by the measurement of the Rossiter-McLaughlin doppler shift by \citet{Winn11}, which indicated that the spin of the primary is aligned as well.

Recently, \citet{Welsh12} reported the discovery of two additional circumbinary gas giants (Kepler-34 b and Kepler-35 b). The relative abundance of these systems among the more than 2,000 eclipsing binaries monitored by Kepler \citep{Slawson11} implies a lower  limit of $\approx 1\%$ in the frequency of circumbinary planets with comparable transit probabilities. Interestingly, all three planets lie just outside the stability boundary for test particles. Their pericenter distance is, respectively, only $\approx$ 6\% (Kepler-34 b), 9\% (Kepler-16 b) and 20\% (Kepler 35-b) larger than the critical semi-major axis, as estimated by the empirical fit in \citet{Holman99}. This represents an important constraint for the formation of the planetary core. Indeed, a natural scenario would entail the planetary core migrating inwards until near the edge of the disk cavity (which will be comparable in extent to the stability boundary for test particles), where the steep gradient of the disk surface density can halt migration \citep{Pierens07}. \citet{Pierens08} simulated the evolution of a 20 $\mass_\s{\earth}$ core, initially placed at the edge of the cavity and free to accrete gas to become a Saturn-mass planet. They found that once the planet depletes the gas in the coorbital region, it will resume a slow inward migration, until its eccentricity is excited and a phase of runaway outward migration is experienced. This runaway migration appeared to stop once the planet crossed the 5:1 resonance with the binary, at which point slow migration is resumed. The ultimate fate of the planet in these simulations is uncertain, due to the long timescales involved. However, it is expected that disk dispersal will ultimately strand the planet on a circular orbit around the binary. Tantalizingly, Kepler-16b lies somewhat close (and outside of) the 5:1 period ratio with the binary.

In this paper, we investigate the conditions for the formation of planetary cores in circumbinary orbits around the Kepler-16 binary system, using a simplified numerical model. We consider the evolution of a disk of $km$-sized planetesimals and determine the impact velocities among planetesimals over $10^5$ years, the typical timescale for runaway and oligarchic accretion \citep{Kokubo00}. These preliminary $N$-body simulations will be used to assess the viability of core accretion as a function of the barycentric semi-major axis.

The plan of the paper is as follows. In \S \ref{sec:setup}, we briefly discuss our numerical model and limitations of our current approach. In \S \ref{sec:drag} we discuss the results of our simulations in the context of planet formation, and conclude in \S \ref{sec:conc}.

\section{Numerical setup}\label{sec:setup}
To conduct our simulations, we use a new hybrid code, {\sc Sphiga} (described in Meschiari et al., 2012, in preparation). {\sc Sphiga} is an $N$-body code that evolves a system of non-interacting test particles (e.g. the planetesimals) subjected to the sum of gravitational forces of massive bodies (e.g. the binary system). In addition, it calculates the frictional force acting on the test particles caused by a putative protoplanetary disk. By default, this is accomplished by following the complete hydrodynamical evolution of the disk with the Smoothed Particle Hydrodynamics scheme   \citep[SPH; see, e.g.,][for recent reviews]{Rosswog09, Price10} in two and three dimensions. The same algorithm used to interpolate the hydrodynamical quantities can be used to interpolate the local gas density and flow and locate possible planetesimal impactors a single loop, leading to significant computational savings. Modelling the self-consistent perturbations from the binary on the disk can alter the planetesimal evolution and potentially increase impact velocities \citep{Marzari08}. Indeed, we expect that non-axisymmetric structure, such as global spiral patterns, will be imposed by the binary, adding a complex time-dependent term. The actual impact of the full hydrodynamical evolution is still uncertain, however. Even bulk quantities such as the disk eccentricity induced by a binary companion appear to depend sensitively on the computational scheme \citep[e.g. the wave damping prescription in][]{Paardekooper08} and the amount of physics modeled \citep[e.g. the equation of state and whether self-gravity was included in][]{Marzari09, Marzari12}.

Nevertheless, significant computational effort is still required to follow the evolution of the combined disk, binary and planetesimal system (with $N_\s{pl} + N_\s{gas} > 10^6$ particles) for at least $\approx 10^5$ binary revolutions. Therefore, for the purpose of this paper, we will use an alternative code path that activates a fixed gas disk. The gas disk exerts a frictional acceleration at the location of the planetesimal given by
\begin{equation}\label{eqn:drag}
\ve{f} = -K |\delta \ve{v}| \delta \ve{v}\ ,
\end{equation}
In Equation \refp{eqn:drag}, $\delta \ve{v} = \ve{v}_\s{pl} - \ve{v}_\s{gas}$ is the relative velocity of the planetesimal with respect to the Keplerian flow of the gas and $K$ is the drag parameter
\begin{equation}
K = \frac{\pi C \rho_\s{g} \radius_\s{pl}^2}{2 \mass_\s{pl}}\ .
\end{equation}
The drag parameter is defined in terms of the planetesimal radius $\radius_\s{pl}$,  the planetesimal mass $\mass_\s{pl}$ (calculated assuming $\rho_\s{pl} = 3$ g/cm$^3$), and the dimensionless coefficient $C$ ($C \approx 0.4$ for spherical bodies). We use the standard prescription of a minimum-mass solar nebula  \citep[MMSN; ][]{Hayashi81} for the disk parameters.  In this configuration, our code and physical setup is functionally equivalent to that used by \citet{Scholl07}. 

To evaluate the collisional speeds among planetesimals, we follow the dynamical evolution of 30,000 test particles uniformly distributed with barycentric semi-major axes between 0.66 and 6 AU; this range includes the current location of the planet ($a_\s{P} \approx 0.7$ AU). The inner boundary was determined by running a simulation with test particles in circular barycentric orbits covering semi-major axes in the range $(1.2 a_\s{b}; 5 a_\s{b})$ for $10^4$ years; we found very good agreement with the fit of \citet{Holman99}. Particles that travel into the inner boundary or become unbound are removed from the simulation. 

The system is initially evolved to $10^5$ years. After this interval, planetesimal-planetesimal close encounters are recorded, with the most important parameter being $\Delta v$, the impact velocity. 
We follow \citet{Fragner11} and \citet{Thebault11} and adopt the prescription for classifying disruptive impacts for planetesimals presented in \citet{Stewart09}. The latter work offers a criterion for catastrophic disruption, the main parameters being the reduced kinetic energy, the masses of the impactors and material properties and constants derived from fits to numerical and laboratory data. Planetesimal collisions are tracked using the inflated radius prescription \citep{Brahic76, Thebault99}, with $\radius_\s{infl} = 5\times 10^{-5}$ AU. The code detects collisions by populating a tree structure at each timestep (as part of the SPH algorithm) and walking the tree to locate the nearest neighbors to each planetesimal with $d < 2\radius_\s{infl}$ \citep[e.g.,][]{Barnes86, Hernquist89}. 

In our simulation, we assign a planetesimal radius for each particle, randomly distributed between 1 and 10 km. We allow for a non-flat primordial distribution in planetesimal sizes by assigning a weight $f(\radius_\s{1}, \radius_\s{2})$ to each impact between planetesimals of radius $\radius_\s{1}$ and $\radius_\s{2}$. Following \citet{Thebault08}, we use a Maxwellian weighting function centered around 5 km with $\sigma = 1$ km. A priori, this choice should yield a more accretion-friendly environment, since it weighs collisions between same-sized planetesimals more than different-sized planetesimals.

\section{Simulations}\label{sec:drag}
\begin{figure}
\epsscale{1.2}
\plotone{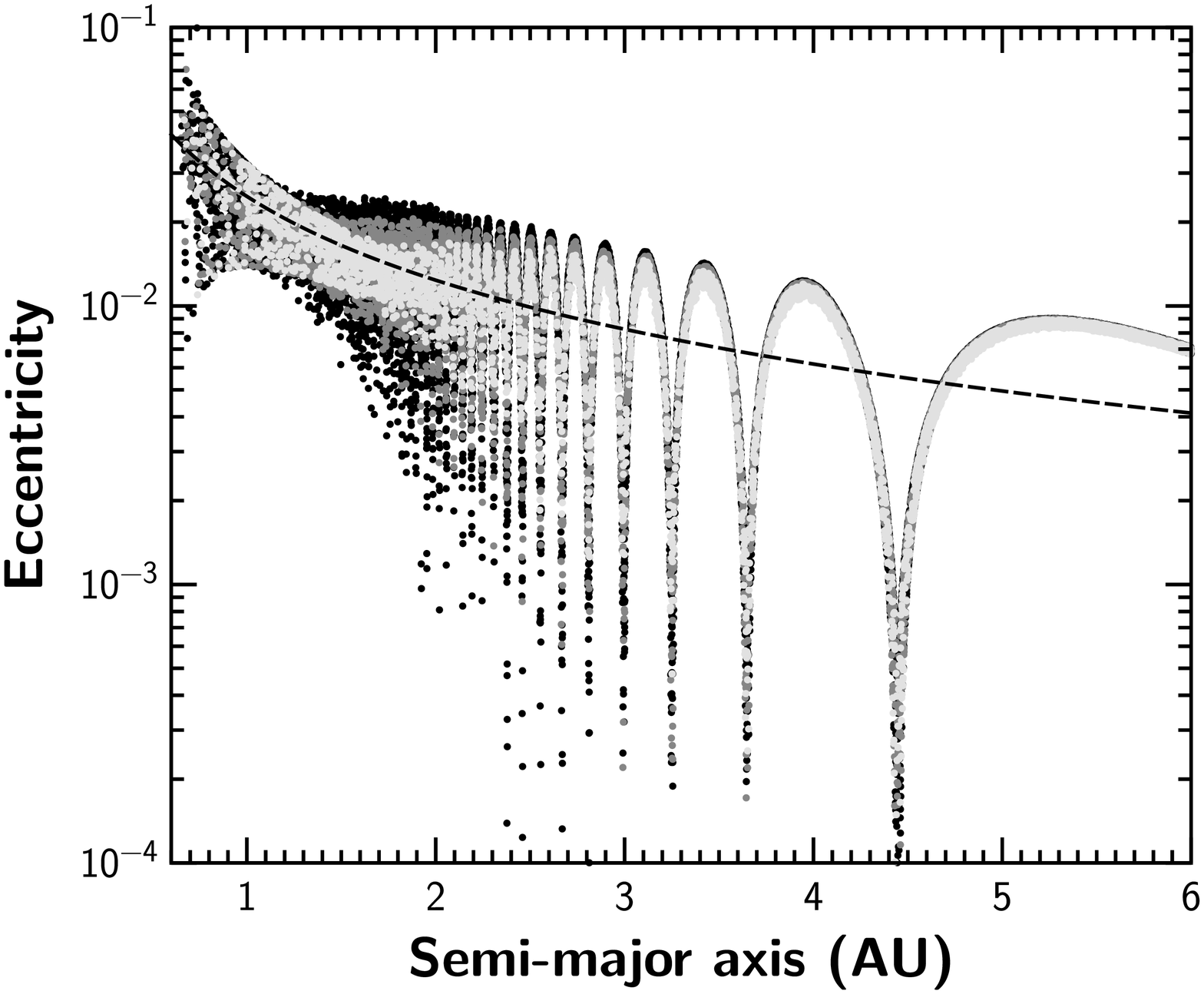}
\plotone{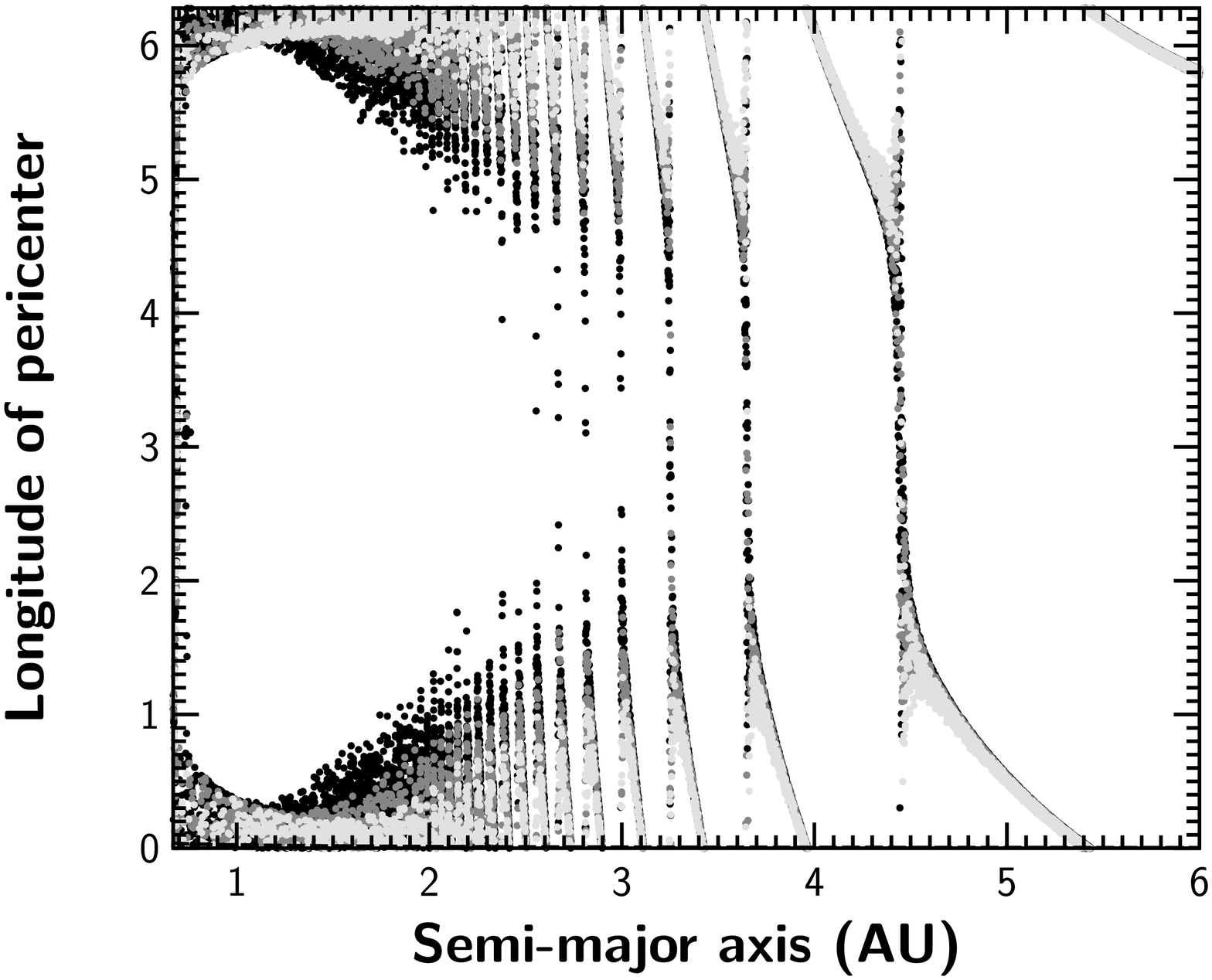}
\caption{Eccentricity $e$ and longitude of pericenter $\varpi - \varpi_\s{B}$ as a function of semi-major axis after $t = 10^5$ years. Planetesimals are colorized with respect to their size: light gray ($1 < \radius_\s{pl} < 4$ km), medium gray ($4 < \radius_\s{pl} < 7$ km), black ($7 < \radius_\s{pl} < 10$ km. The dashed line shows the forced eccentricity.}\label{fig:ecc_gas}
\end{figure}

\begin{figure}
\epsscale{1.2}
\plotone{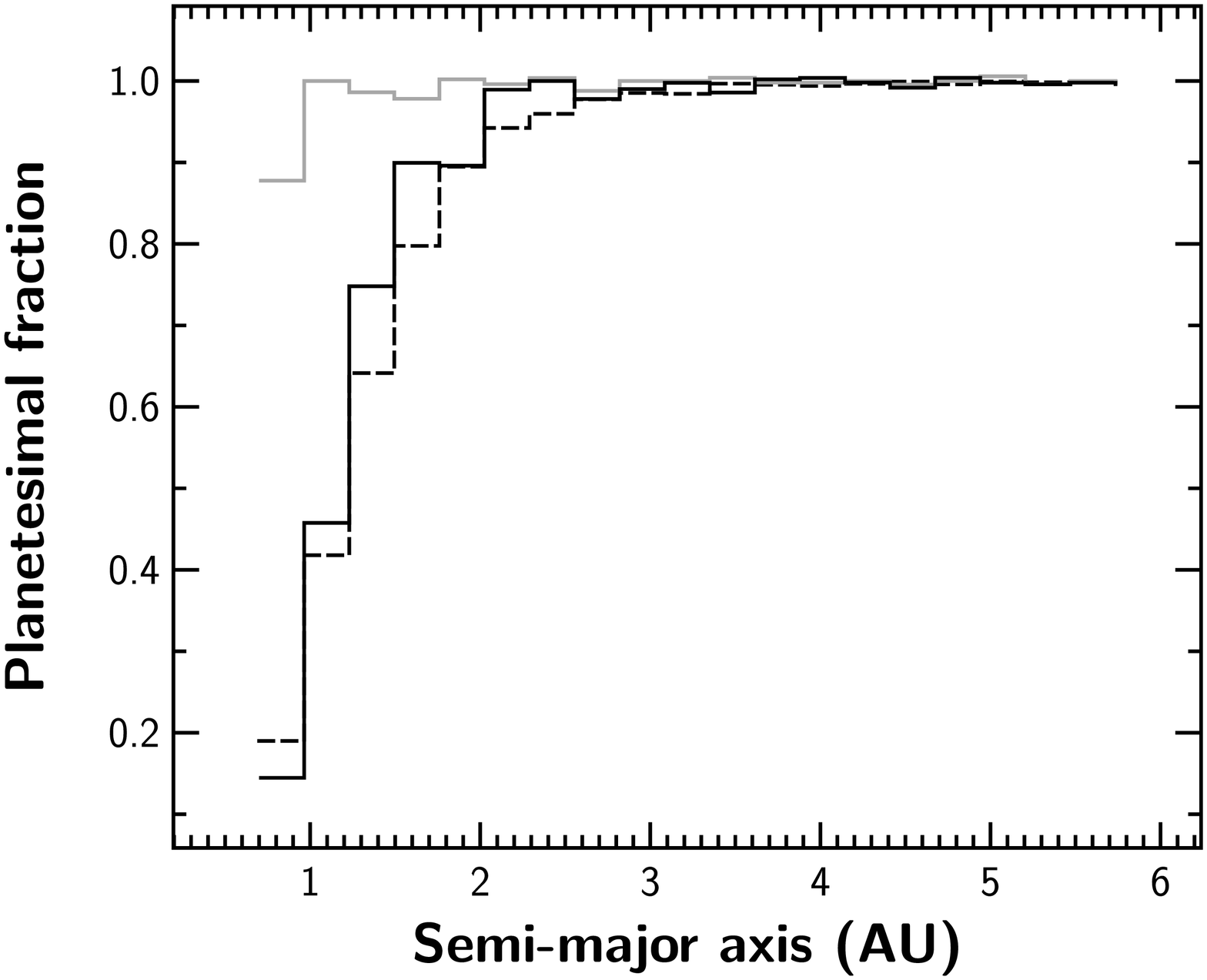}
\plotone{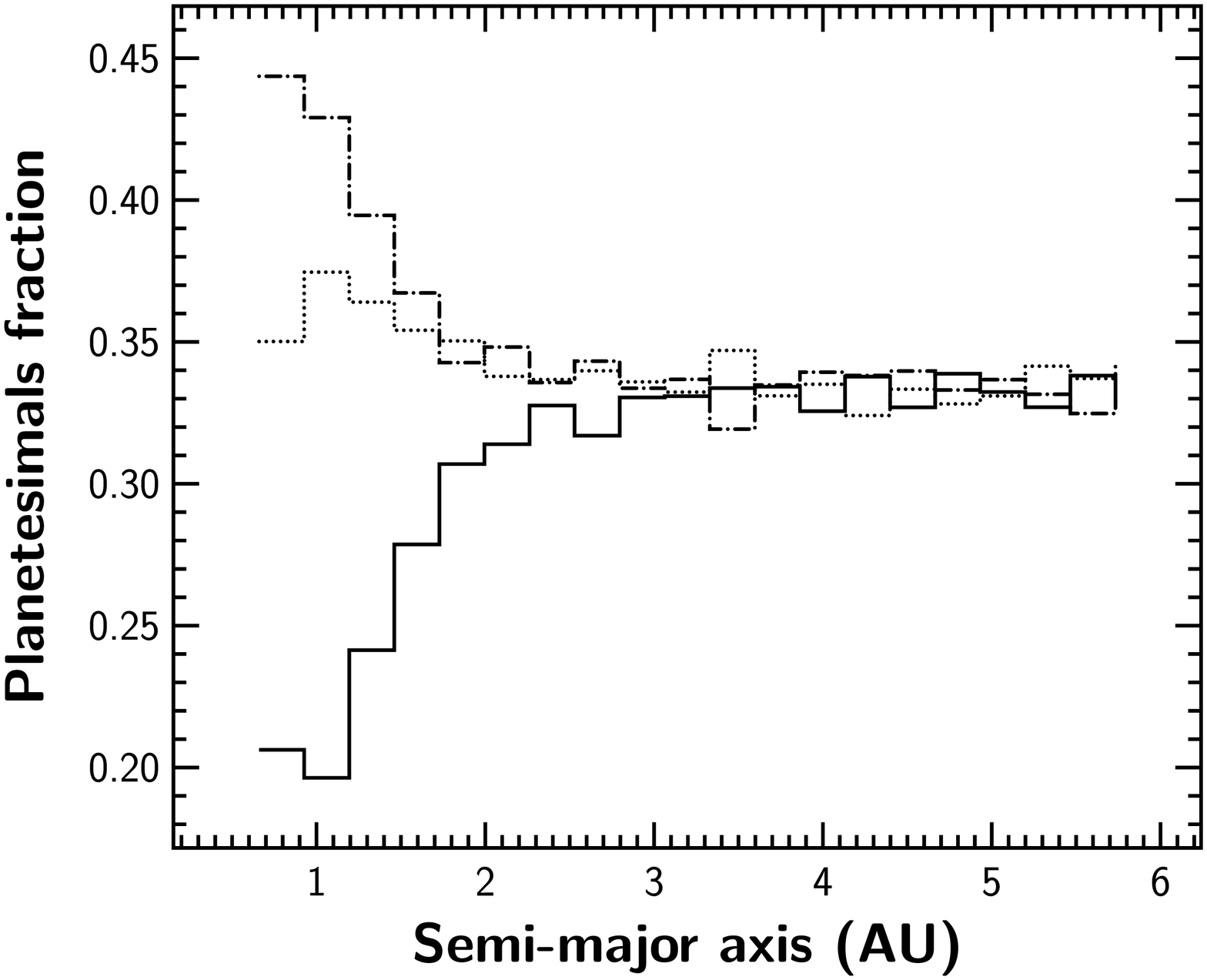}
\caption{\emph{(Top)} Planetesimal number binned in semi-major axis after $t = 10^5$ years, normalized by the initial distribution in semi-major axis. The $N$-body run (black line) and the output from an analytic model for the single star case (gray line) and with a forced eccentricity term (dashed line) are shown. \emph{(Bottom)} Relative fractions of planetesimals with $1 < \radius_\s{pl} < 4$ km (solid line), $4 < \radius_\s{pl} < 7$ km (dotted line) and $7 < \radius_\s{pl} < 10$ km (dash-dotted line) after $10^5$ years.}\label{fig:drift}
\end{figure}

\begin{figure}
\epsscale{1.2}
\plotone{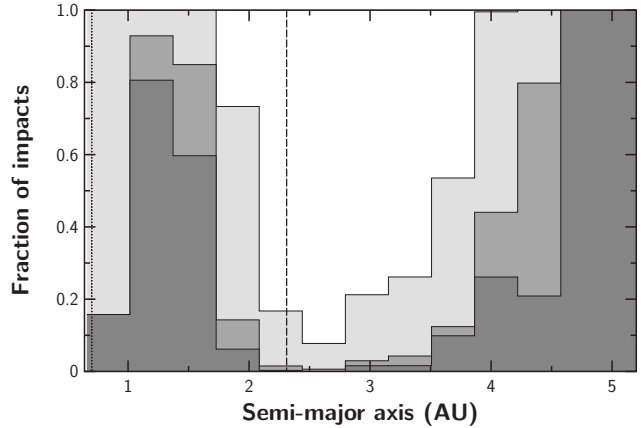}
\caption{Fraction of accreting (dark gray), disturbed (medium gray), uncertain (light gray) and erosive impacts (white), as a function of semi-major axis, after $t = 10^5$ years. The present-day location of the planet and the fiducial ice line are plotted (dotted and dashed line, respectively).}\label{fig:enc_gas}
\end{figure}
As expected, the planetesimals are quickly perturbed from their initial low-eccentricity configuration by the gravitational stirring of the central binary. Their eccentricities initially oscillate around the forced eccentricity
\begin{equation}
e_\s{f} = \frac{5}{4}(1 - 2\mu) \frac{a_\s{B}}{a} e_\s{B}
\label{eqn:forcedecc}\end{equation}
\citep{Moriwaki04}.  The presence of gas drag tends to damp the eccentricity oscillations towards the forced eccentricity over longer timescales. Damping and periastron phasing will be more effective for smaller (since the gas drag coefficient is proportional to $\radius_\s{pl}^{-1}$) and close-in bodies (since $\rho_{gas} \propto a^{-2.75}$). However, the eccentricity spread remains somewhat large at small semi-major axes, where the gravitational perturbation of the central binary acts to pump eccentricities. At large semi-major axes, where the damping timescale is longer, the values of eccentricity tend to their counterparts in gas-free simulations.

In the inner parts of the disk, planetesimals will spiral into the inner boundary due to radial drift. The radial drift timescale can be estimated by assuming the planetesimal loses angular momentum slowly due to the torque from the headwind of the gas \citep{Weidenschilling77}. For the drag prescription of Equation \ref{eqn:drag}, we find an estimate for the infall timescale (in units where $G\mass = 1$) is given by
\begin{equation}
\tau_\s{rd} = \frac{a_\s{pl}}{v_\s{rd}} \approx \frac{4}{3} C^{-1} \mass_* \frac{\rho_\s{pl}}{\rho_\s{gas}} \frac{\radius_\s{pl}}{a_\s{pl}^{1/2}} \left(\delta v\right)^{-2}\ ,
\label{eqn:drift}\end{equation}
where $\rho_\s{pl}$ is the density of the planetesimal and $\mass_*$ is the total mass of the binary. 

In the case of planet formation around single stars, eccentricities are very low and $\delta v \sim h_\s{0}^2 v_\s{kep}$ is mainly determined by the local scale height $h_\s{0}$, with a typical drift timescale at 1 AU of $10^6$ years for a 5-km planetesimal. In the circumbinary environment, on the other hand, the perturbation from the binary companion acts to raise eccentricities throughout the planetesimal disk, such that the dominant term contributing to $\delta v$ is given by the time-varying speed of the planetesimal sampling different gas velocities at the apsides. 

Figure \ref{fig:drift} shows the distribution of planetesimals after $t = 10^5$ years, binned in semi-major axis. We find that inside $\approx 1.5$ AU, the planetesimal disk is severely depleted. Indeed, in our setup, drift timescales are a strong function of semi-major axis ($\propto a^{-5/2}$), such that radial drift from the outer parts of the disk cannot replenish the inner disk effectively. We compared the planetesimal distribution of our $N$-body run with an analytic model based on Equation \ref{eqn:forcedecc} and \ref{eqn:drift}. Assuming $\delta v \approx 0.5 e_\s{f} v_\s{kep}$, we find good agreement between the two. Finally, the second panel of Figure \ref{fig:drift} shows that the distribution of planetesimal sizes is skewed towards larger planetesimals at small semi-major axes, since larger planetesimals are less affected by the gas drag. This can contribute to making the inner region more accretion-friendly for two reasons. Firstly, larger planetesimals can withstand larger impact velocities. Secondly, the spread in sizes will be reduced, which means that the spread in the phasing of the planetesimals will also be reduced.

In the outer parts of the disk, where damping is less effective, planetesimals are initially weakly phased because the oscillations are coherent and spatially extended; therefore, impact velocities tend to be lower. However, the frequency of the oscillation around the forced eccentricity increases with time, ultimately leading to orbital crossing \citep{Thebault06}. The orbital crossing boundary $a_\s{cross}$ sweeps outwards in semi-major axis, increasing impact velocities. In our simulation, collisions are recorded for a small time window after $t = 10^5$ years. As evidenced in Figure \ref{fig:ecc_gas}, regions outside $\approx$ 3.5 AU ($\approx 13 a_\s{B}$) have not experienced orbital crossing. This is expected, since $a_\s{cross}$ is a weak function of time \citep{Thebault06}. Over longer timescales, the impact velocities will increase in the outer regions, as they are swept by the orbital crossing boundary. However, we expect the core of Kepler-16 b to be formed and accreting gas before significant gas dispersal has occurred  \citep[on a timescale of $\approx 10^5$ years; e.g.][]{Wolk96}.

\subsection{Implications for planet formation}
Figure \ref{fig:enc_gas} shows the fraction of accreting encounters as a function of semi-major axis. We found that the following qualitative situation holds for different radial locations:
\begin{enumerate}[(a)]
\item in the region between the stability boundary and 1 AU (which includes the present-day location of the planet $a_P \approx 0.7$ AU), eccentricities are pumped to high values by the central binary and planetesimal number density is low due to the fast radial drift. The majority of encounters are in the ``uncertain'' regime, with the potential of being accreting depending on the prescription for the outcome of disruptive collisions.
\item for a small range in semi-major axis outside 1 AU, the spread in $e$ and $\varpi$ is smaller and planetesimal distributions are skewed towards larger planetesimals. The majority of encounters are accreting.
\item between 1.75 AU and 4 AUs, the magnitude of the eccentricity and the differential phasing raises the impact velocities, such that the majority of the encounters are erosive.
\item outside 4 AUs, orbital crossing has not been realized yet and gas drag is weaker due to the steep radial dependence of the gas density; therefore, orbits are only weakly phased. The majority of encounters are accreting.
\end{enumerate}

We conclude that planet formation is likely inhibited for a large range in semi-major axis (location (c), between 1.75 and 4 AUs). This range in semi-major axis includes the nominal location of the ice line for an irradiated disk, estimated from the scaling $a_\s{ice} \sim 2.7 \mathrm{AU}\ (\mass/\mass_\s{\odot})^2 \approx 2.3$ AU \citep{IdaLin04}, assuming $\mass = \mass_\s{A} + \mass_\s{B}$. 

What is the impact of this ``forbidden region'' for planet formation? It is instructive to refer to the predictions of the standard core accretion paradigm for single stars; in particular, the outcome of large-scale Monte-Carlo planet synthesis models \citep[e.g.][]{IdaLin04, Mordasini09}. \citet{Mordasini12} recently conducted a Monte-Carlo planet synthesis simulation for a variety of disk masses and metallicities, for the nominal case of a 1 $\mass_\s{\odot}$ central star. In the core accretion paradigm, metallicity represents a threshold quantity for the formation of planetary cores. Accordingly, they found that the cores of giant planets ($\mass \gtrsim \mass_\s{J}$) tend to preferentially form outside the ice line when the metallicity (which acts as a proxy for the solid content of the disk) is low. The actual location of the ice line scales with the disk mass, which contributes to the spread in  semi-major axis.

\begin{figure}
\epsscale{1.2}
\plotone{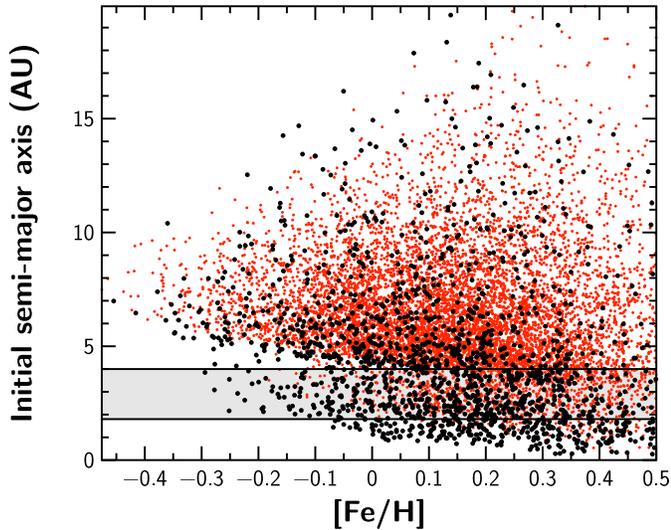}
\caption{Initial location of embryos that grow to final masses $0.2 < \mass_\s{P} < 0.4 \mass_\s{J}$ (black points) and $\mass_\s{final} > 1 \mass_\s{J}$ (red dots) in the simulations of \citet{Mordasini12}, for a range of metallicities and disk masses. The shaded region corresponds to the range in semi-major axis where embryo formation is disturbed in the Kepler-16 system.}\label{fig:pop}
\end{figure}

In Figure \refp{fig:pop}, we plot a different subset of the output of the simulations of \citet{Mordasini12}\footnote{http://www.mpia-hd.mpg.de/homes/mordasini/Site7.html}, focusing on the ensemble of embryos that acquire masses comparable to Kepler-16 b ($0.2 \mass_\s{J} < \mass < 0.4 \mass_\s{J}$). The initial location of the embryo is plotted as a function of metallicity. For disks of solar or super-solar metallicity, such planets are formed throughout the disk, with a substantial fraction formed inside 2 AU (about 40\%). At subsolar metallicities comparable to Kepler-16 ([Fe/H] $\approx -0.3 \pm 0.2$), however, such cores are only found outside 2 AU, with a minority lying in location (c) (about 20\%). While the synthetic population refers to the nominal 1 $\mass_\s{\odot}$ single star case, with one embryo per disk, it suggests that \textit{in situ} planet formation in location (a) might be hampered by the low surface density in solids at 1 AU. Our simulations place an additional dynamical constraint, indicating that less than 20\% of encounters within 1 AU are accreting. This, compounded with the low planetesimal density in the region (Figure \ref{fig:drift}), makes \textit{in situ} formation of a substantial core difficult.

Finally, it is also crucial to recognize that non-axisymmetric perturbations from the disk might play an important role in the dynamics of the inner disk. The eccentric central binary will likely excite spiral structures, which might act to pump the eccentricity of the inner planetesimals and alter the phasing of their orbits. Indeed, \citet{Marzari08} conducted full 2D hydrodynamical simulations with a small number of tracer planetesimals embedded in the disk, and found significant oscillations in the eccentricity and longitude of pericenter around the equilibrium value.

\section{Discussion}\label{sec:conc}
Planet formation in presence of close binaries presents a number of challenges to the traditional core accretion paradigm. Historically, most of the theoretical effort has been expended to study pathways to planet formation in S-type orbits for planets that had been observed through RV surveys, or targets with observationally desirable properties (e.g., $\alpha$ Centauri). With the launch of \Kepler{}, however, we expect that the sample of planets in P-type orbits around eclipsing binaries will rapidly outnumber the handful of planets in circumstellar configurations detected with RV surveys. Indeed, a sample of 750 \Kepler{} targets are eclipsing binaries for which eclipses of both stars are observed, and a subset of 18\% exhibited deviations in the timing of the eclipses \citep{Welsh12}. Since the definitive determination of the planetary nature of a putative KOI relies on the detection of tertiary and quaternary eclipses, we expect that as the baseline of the observation increases, more KOIs will be confirmed as genuine circumbinary objects. 

In this paper, we have conducted a preliminary simulation of the feasibility of circumbinary planet formation in the Kepler-16 system. In accordance to an earlier study conducted by \citet{Scholl07} for a different set of binary parameters, we have found that, for generous initial conditions that favor planetesimal accretion, planet formation appears to be feasible far enough from the central binary. However, we have identified a substantial radial span between 1.75 and 4 AU where planet formation is strongly inhibited. Within the planet accretion framework, the most likely sequence of event is the formation of a core outside the forbidden region, followed by inwards migration driven by tidal interaction with the protoplanetary disk \citep{Pierens07}. Although we measured impact velocities potentially favorable to accretion close to the present-day location of the planet, \textit{in situ} formation of Kepler-16 b is less likely due to overall high encounter speeds, low planetesimal density, low metallicity of the star, and non-axisymmetric perturbations from the disk (not modeled in this simulation). 

We remark that the simulations presented in this paper only demonstrate that, \textit{choosing the most favorable conditions for planetesimal accretion} and an assumed initial planetesimal size of 1-10 km, the formation of an embryo outside 4 AU is plausible, with traditional migration processes subsequently moving the planet to its current location. Our approach has several limitations introduced for the sake of simplicity and computational speed; chiefly, we disregarded the evolution of the protoplanetary disk and the collisional outcome of planetesimal impacts. For the former, we plan to follow approximately the hydrodynamical response of the disk with the SPH algorithm included in the \SPHIGA{} code in a follow-up paper. For the latter, a time-dependent distribution of planetesimal sizes would more accurately model the extent of the accretion-friendly regions, which depend sensitively on the planetesimal parameters. The numerical procedure of \citet{Paardekooper10} represents a possible approach to following the collisional evolution of the planetesimal size distribution. 

\acknowledgments
We acknowledge support from the NASA Grant NNX11A145A. The author thanks Greg Laughlin and the referee, Philippe Thebault, for useful discussions.
\\
\phantom{}
\bibliographystyle{apj}

\end{document}